\documentclass{ws-procs975x65}  
\usepackage{graphicx}
\usepackage[]{natbib}
\usepackage[flushleft]{threeparttable}

\begin{document}
\title{The local effect of Dark Energy in galaxy clusters}

\author{Martina Donnari$^1$, Marco Merafina and Manuel Arca-Sedda}

\address{Department of Physics, Sapienza University of Rome,\\
Rome, 00100, Italy\\
$^1$E-mail: martina.donnari@uniroma1.it}

\begin{abstract}
Recently, observational data and high precision mapping of the local velocity field of Local Group and Virgo cluster have revealed a linear velocity-distance relation of the outermost galaxies, properly referred to as Local Hubble Flow.
By means of direct N-body method, we performed several simulations in which a galaxy cluster undergoes the action of the Dark Energy force and of the gravitational one induced by the gas. We reproduced the so-called Hubble diagrams, to highlight the outflow of the galaxies lying in the external region of the cluster. Our preliminary results suggest that the observed outflow of galaxies is likely due to the local effect of Dark Energy. Furthermore, the accuracy of the N-body method used, allows us to follow the merging process among some galaxies with the aim to reproduce the formation of a single compact object in the centre of the cluster.
\end{abstract}


\bodymatter

\section{INTRODUCTION}
According to the $\Lambda$ Cold Dark Matter paradigm ($\Lambda$CDM), all celestial bodies are embedded in a perfectly uniform Dark Energy (DE) background, which generate a repulsive force opposite to the gravitational one. 
Recent observational works have revealed the presence of the so-called \textit{Local Hubble Flow} (LHF), a regular velocity-distance relation of galaxies that lie in the outermost region of the clusters, seen in our Local Group and in the nearest Virgo cluster \citep{2002Kara,2015Chernin}.

Through numerical simulations made by a direct N-body code, this work aims to quantify the role of the DE on the flow out of galaxies from the cluster centre, by investigating if DE can have strong dynamical effects on \textit{small} cosmic scales, i.e. a single galaxy cluster.

\section{MODELS OF SIMULATIONS}
We show the dynamical evolution of a single galaxy cluster composed by a total number of particles greater than $10^6$, divided into $N_g=240$ galaxies. 
It is trivial to stress that a real cluster of galaxies contains a number of stars in the range $10^{13}-10^{15}$, but such small number is enough for a sufficient precise description of the global dynamics within the cluster.

\subsection{Modeling the galaxy cluster}
The distribution of the galaxies masses in the cluster is \citep{2007tutukov} $f(M_g)=k \, M_g^{-1}$. 
We selected $M_g$ in the range $(9\times 10^{10} - 10^{12})$ M$_\odot$.
Each galaxy have been modeled with the so-called Dehnen $\gamma$-models \citep{1993Dehnen}, whose density profiles are given by:
\begin{equation}
\rho(r)=\frac{(3-\gamma)M_g}{4\pi r_g^3}\left(\frac{r}{r_g}\right)^{-\gamma}\left(1+\frac{r}{r_g}\right)^{-4+\gamma},
\label{dehnen}
\end{equation}
where $M_g$ is the total mass of the galaxy, $r_g$ its lenght scale and $\gamma$ is the inner slope of the profile.
The value of $r_g$ is obtained by using the following relation \citep{2014mas}:

\begin{equation}
r_g(\mathrm{kpc}) = 2.37 \, (2^{1/(3-\gamma)}-1)\left(\frac{M_g}{10^{11}M_\odot}\right)^{0.14}.
\end{equation}
Moreover, the $\gamma$ values are randomly assigned to each galaxy in the range $0.2-1.75$.
At the end, we gathered a sample of 240.
The total mass of the cluster is $M_{cl} = 9.2\times 10^{13}$M$_\odot$.
We distributed the galaxies in the space using a King-like profile \cite{1998girardi}, due to 

\begin{equation}
\rho_{cl}(r)=\rho_0\frac{1}{(1+(r/r_c)^2)^{\alpha}}\frac{1}{{\rm cosh}(r/r_{cut})},
\end{equation}
being $\rho_0$ the central density, $r_c=0.1$ Mpc the core radius, $r_{cut}=3.85$ Mpc the truncation radius and $\alpha=1$ that corresponds to the classical King distribution. 

\subsection{Modeling the external potentials}
\label{ModelingDE}
According to $\Lambda$CDM model, we assumed $\rho_{\Lambda} = 0.7\times10^{-29} \rm{g/cm^3}$. When DE is taken into account, each particle moves under an additional acceleration arising from antigravitational effect of DE.
The interplay between gravity and anitgravity leads to define a distance $R_{\Lambda}$ at which the total force is null, i.e.

\begin{equation}
R_{\Lambda}=\left(\frac{3M_{cl}}{8\pi \rho_{\Lambda}}\right)^{1/3} \; .
\label{Rlambda}
\end{equation}
This physical quantity is the so-called \textit{zero-gravity radius} (ZGR) \citep{2003Chernin,2001Chernin,2008Chernin}. The gravity dominates at distances $r<R_{\Lambda}$, whereas the antigravity is stronger that the gravitational force at distances $r>R_{\Lambda}$.

From X-ray astronomy, we know that the volumes of space between galaxies in galaxy clusters are filled with a hot plasma ($10^7-10^8$ K), better known as Intra-Cluster Medium (ICM). Therefore, in order to obtain a more reliable model, we simulated gravitational contribution of the ICM with another external potential given by the modified $\beta$-model \citep{1976cavaliere,2006Vikhlinin} which describe the spatial distribution of the ICM in the majority of the observed clusters. The density profile for this model is given by

\begin{equation}
\rho_{gas}(r)=\rho_0\left(\frac{r}{r_c}\right)^{-\alpha/2}\left[1+\left(\frac{r}{r_c}\right)^2\right]^{-3\beta/2+\alpha/4} \; .
\label{beta-model}
\end{equation}
The value of the parameters are $\alpha=4.6$ and $\beta=1.2$ and the gas core radius is $r_c=0.25$ Mpc.  
Hence, each particles suffers a total forces given by 
\begin{equation}
F_{\rm tot}=F_{\rm grav}+F_{\rm DE}+F_{\rm gas} \; .
\end{equation}
When the ICM is taken into account, an additional mass, equal to $\sim$ 5$\%$ of the total mass of the cluster, is considered ($M_{gas}\simeq 4.7\times 10^{12} \, \rm{M_{\odot}}$).

\subsection{Set of simulations}
We performed four simulations, listed in the following and resumed in Tab. \ref{tbl1}
\begin{itemize}
\item{S1}: the cluster is isolated in the space, no external forces act on it;
\item{S2}: the cluster is embedded in a Universe filled with DE;
\item{S3}: the cluster is embedded in an external potential simulating the ICM;
\item{S4}: the cluster is embedded in both potentials due to DE and ICM.
\end{itemize}

\begin{table}
\tbl{Galaxy cluster models}
{\begin{tabular}{@{}ccccc@{}}
\toprule
Model & $\rho_{\Lambda}$ & M$_{\rm gas}$ & T$_{\rm ev}$ \\
      & ($10^{-30} \; \rm g \, cm^{-3}$) & ($10^{12}$ \, M$_{\odot}$) & (Gyr)\\
\hline
S1 & 0 & 0 & 3.1 \\
S2 & 7 & 0 & 3.7 \\
S3 & 0 & 4.7 & 3.5 \\
S4 & 7 & 4.7 & 5.3 \\
\botrule
\end{tabular}}
\begin{tabnote}
Components enabled in each model and simulated time.\\
\end{tabnote}
\label{tbl1}
\end{table}

Because of the huge computational cost, the models S1, S2 and S3 were carried out up to $\sim$ 3.5 Gyr whereas the model S4 was carried out until 5.3 Gyr which is enough to detect well visible differences among all the four cases.

\section{ANALYSIS AND RESULTS}
For each simulations we investigate the motion of the galaxies in the further region of the cluster, close to the ZGR. It has been also reproduced the Hubble diagram which represents the velocity-distance relation of each galaxies in order to quantify the effects of DE in the galactic dynamics. Moreover, we followed the formation of a merger product in the central region of the cluster. All the results are discussed in the following sections.

\subsection{Dynamics of the further galaxies}
As it was mentioned in Section \ref{ModelingDE}, $R_{\Lambda}$ is the distance at which the forces that act on a galaxy cluster belance each other. 
In order to evaluate the effect of DE antigravitational force, we investigated the trajectory of the outermost galaxies during all the four simulations, founding interesting results in the models in which the DE is switched on. 
In order to label one galaxy like an {\itshape{escaper galaxy}}, it must be at distance greater than the ZGR ($r>R_{\Lambda}$) and it must have a positive total energy ($E_{tot}>0$).
When this two criteria are satisfied, that galaxy can be considered unbound to the cluster.
\begin{figure}[!ht]
\begin{center}
\includegraphics[width=2in]{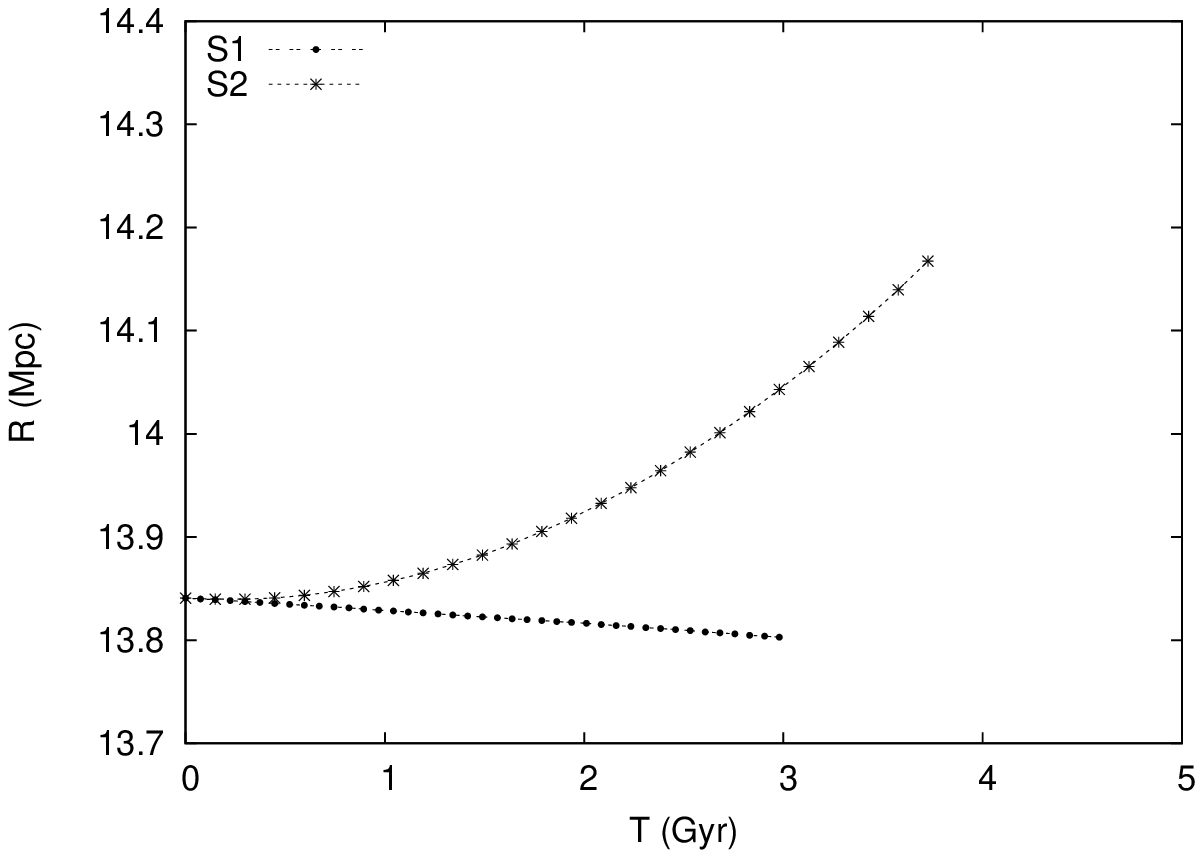}
\includegraphics[width=2in]{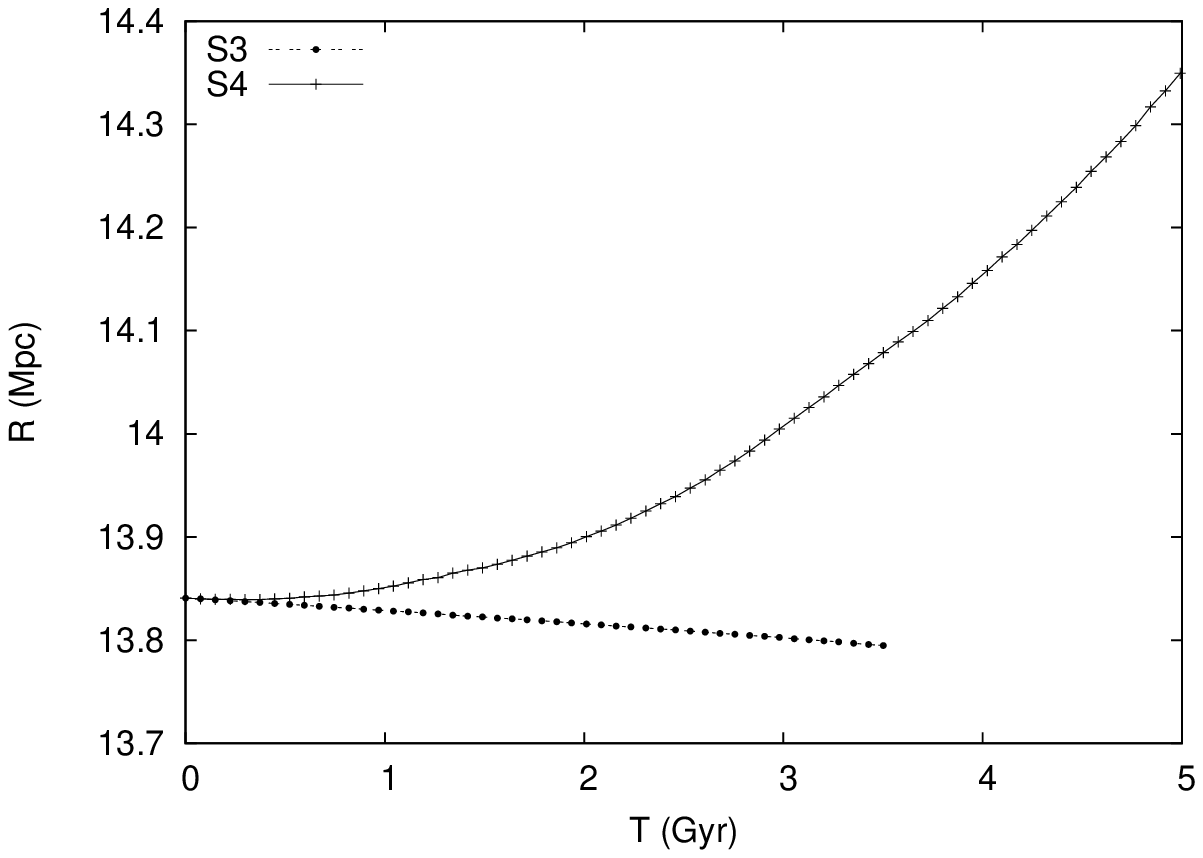}
\end{center}
\caption{Trajectory of the CoM of one galaxy with $E_{tor}>0$ and placed at distance $r>R_{\Lambda}$. \textit{Left}: models S1 and S2; \textit{right}: models S3 and S4.}
\label{Fig.1}
\end{figure}
In Fig. \ref{Fig.1} is well visible that for models S2 and S4, in which the DE is considered, the galaxy go away from the cluster centre, whereas in simulations S1 and S3 the attractive effect of the gravity and the gas, moves the galaxy toward the cluster centre. This picture is reproduced by $\sim$ 90$\%$ of the escaper galaxies, providing an evidence for the dynamical effects of DE also on scales of few megaparsecs, as well as in larger volumes.

\subsection{Hubble diagrams}
A way to higlight the effect of DE is looking at the so-called Hubble diagrams, which show the radial velocity of each galaxy with respect to the distance from the cluster centre. 

On the largest scales, i.e. $>100$ Mpc, the Universe is homogeneous and isotropic and well described by the Friedmann models. In this context, the Hubble constant $H_0$ is the cosmological parameter that well describe the size and the age of the Universe.
On the other hand, on small scales ($<100$ Mpc) the Universe is significantly inhomogeneous, due to the presence of galaxies and cluster of galaxies. Because of density fluctuations, we are in presence of a different expansion rate that can be called \textit{local} Hubble constant $H_L$, different from the global one.

These Hubble diagrams can be used to make a roughly estimation of $H_L$, which represent the slope of the velocity-distance relation.
\begin{figure}[!ht]
\begin{center}
\includegraphics[width=2in]{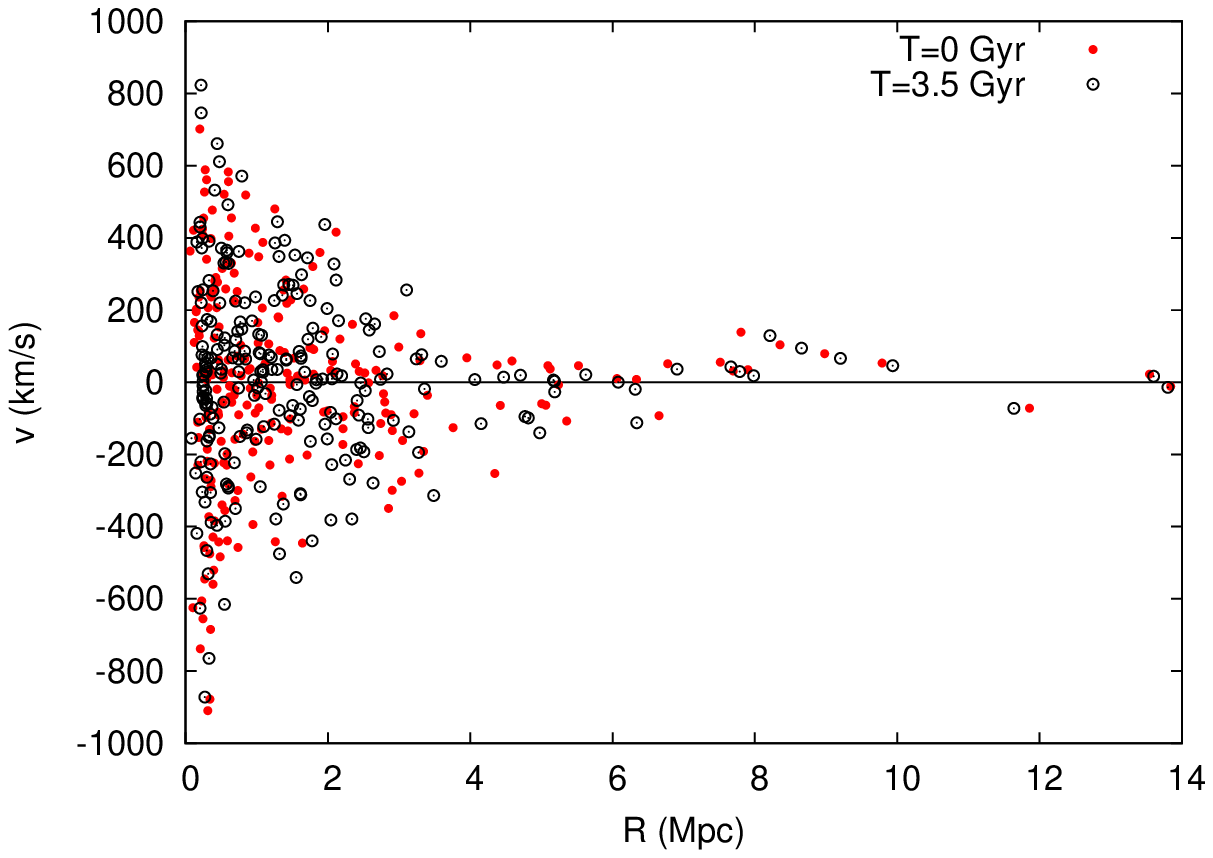}
\includegraphics[width=2in]{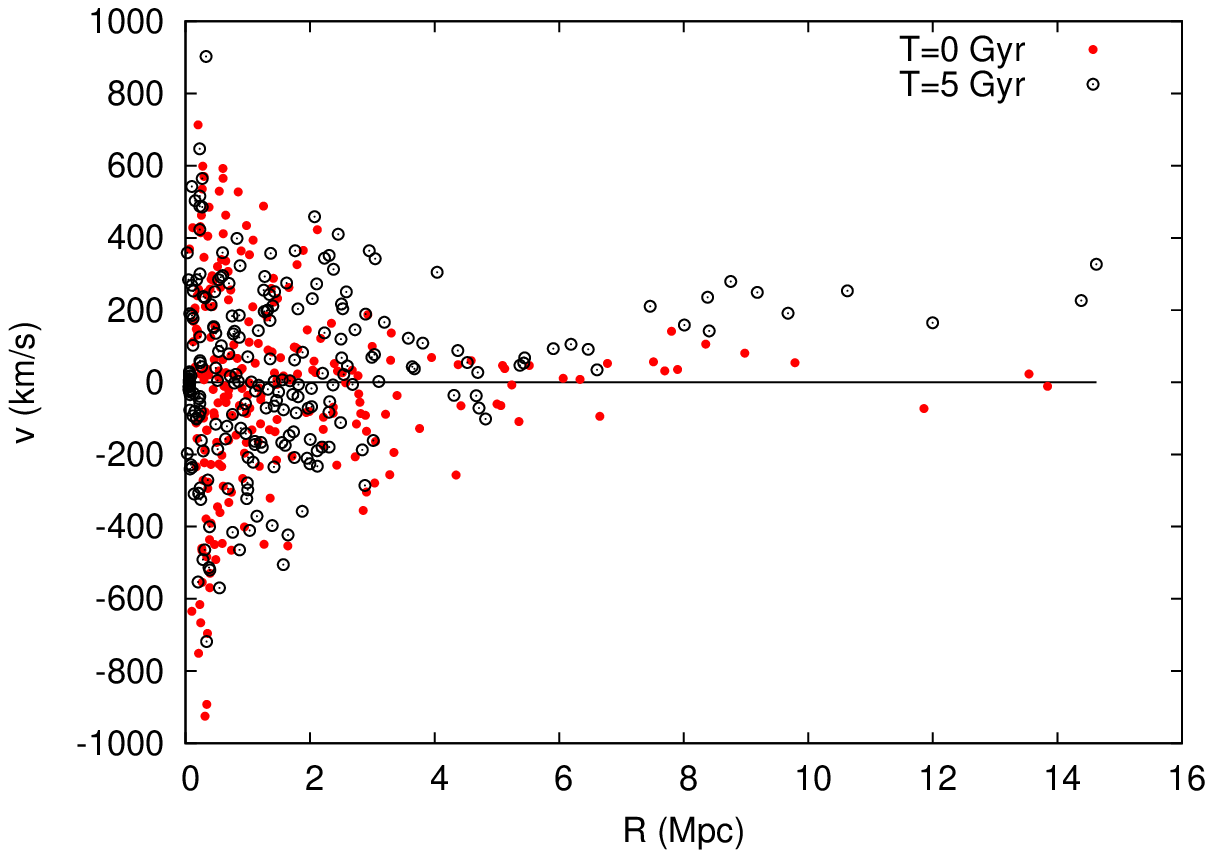}
\end{center}
\caption{Hubble diagrams of the simulated galaxy cluster in S1 (left panel) and S4 (right panel).}
\label{Fig.2}
\end{figure}
In the left panel of Fig. \ref{Fig.2} are represented the radial velocities of the 240 galaxies in S1 from the starting point (dots) to the end of the simulation (circles), at 3.5 Gyr. It is possible to see that the average of the radial velocity is very close to zero. On the other hand, in the right panel of Fig. \ref{Fig.2} are represented the galaxies velocities in S4 from the starting point (dots) up to $\sim$ 5.3 Gyr of the evolution (circles). Here is evident a nearly linear increase of the velocity at increasing distance. This results suggest that there are no possibility to have a LHF if DE is not taken into account.

\subsection{Cluster substructures}
\begin{figure}
\centering
\includegraphics[width=2in]{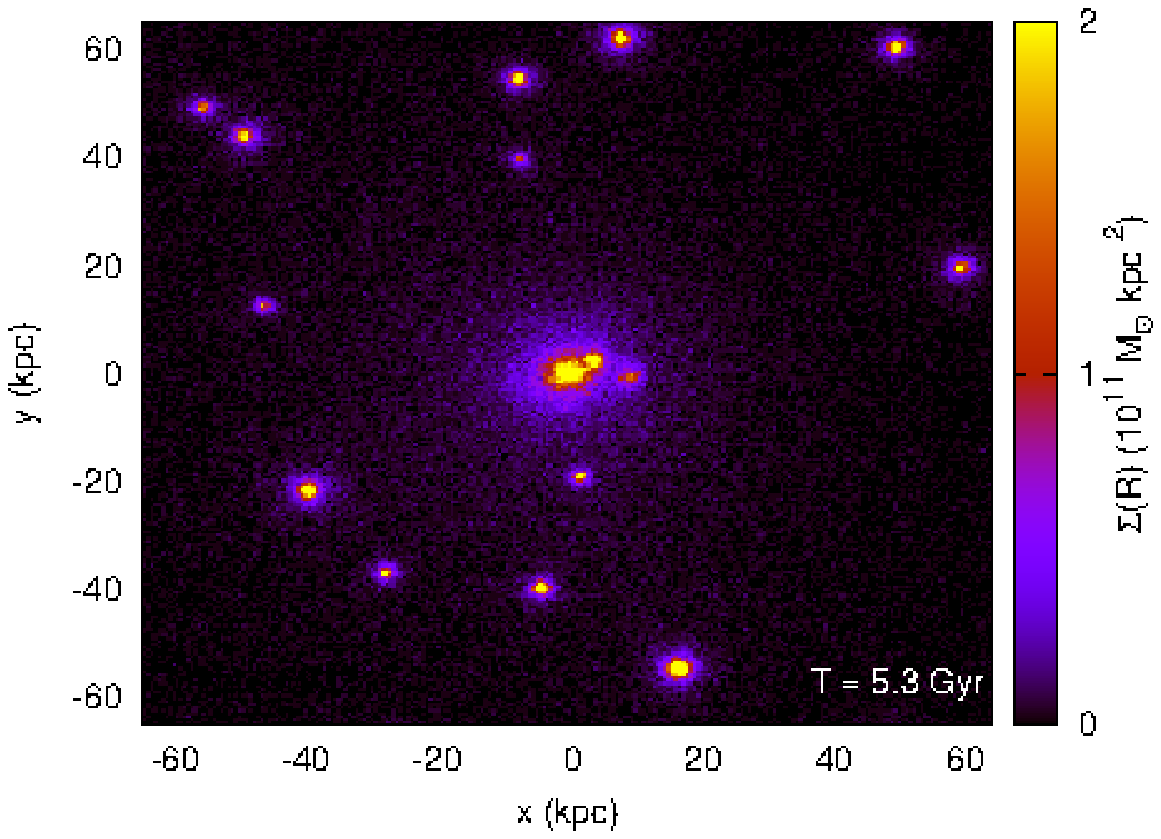}
\includegraphics[width=2in]{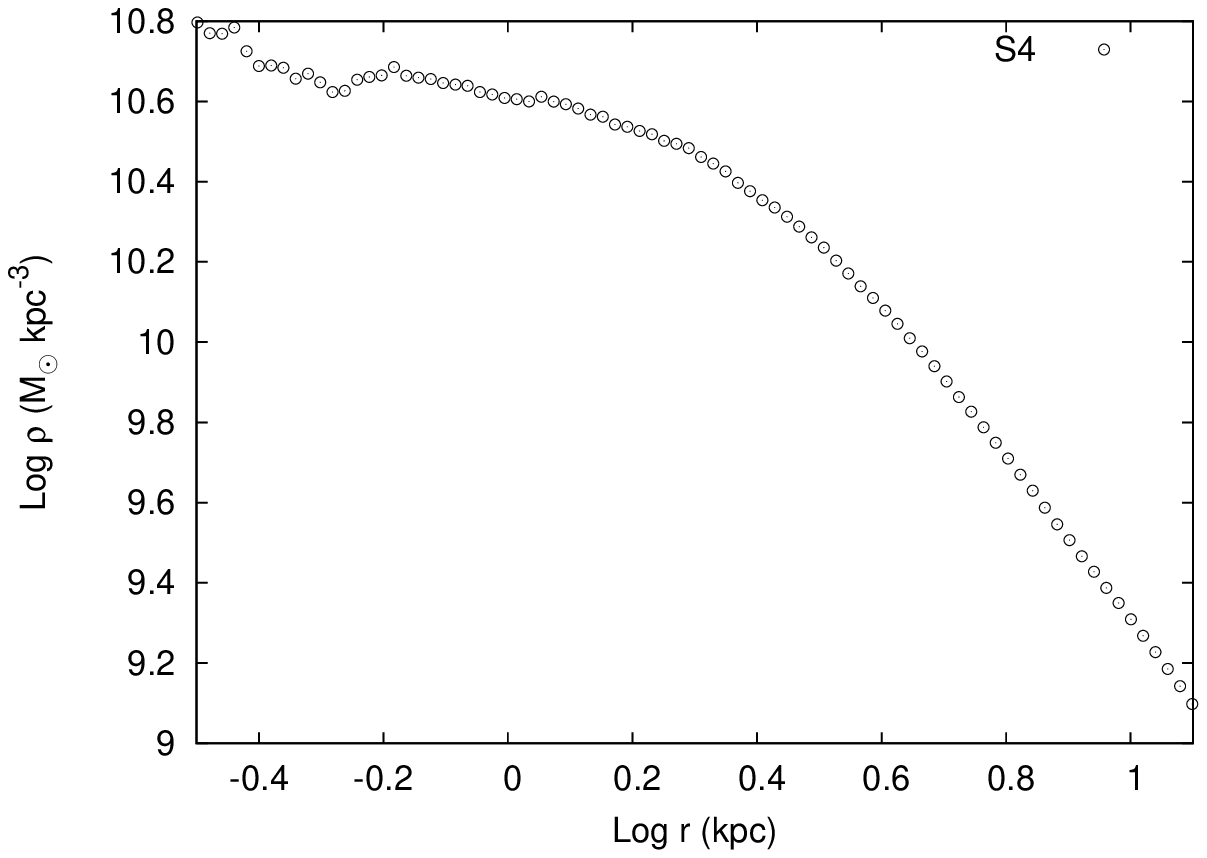}
\caption{\textit{Left}: Surface density map of the GC nucleus in model S4 after $5.3$ Gyr. \textit{Right}: Density profile of the merging product.}
\label{map}
\end{figure}

We found that in all the four cases the most massive galaxies tend to concentrate to the cluster centre, due to the action of dynamical friction. Their collisions and merging drive the formation of  massive substructures.
Our analysis was focused on S4 model. The mass of the structure is $M_{\rm cen}=1.2 \times 10^{12}$ M$_\odot$, enclosed within $10$ kpc from the GC centre. The left panel of Fig. \ref{map} shows a surface density map of the inner $70$ kpc of model S4. It is evident the central structure, surrounded by a number of smaller galaxies which are approaching it. 
We found that the density profile (right panel of Fig. \ref{map}) of the merging product is well described by a Dehnen profile due to Eq. (\ref{dehnen}), where $M_g = (3.27\pm 0.05)\times 10^{12}$ M$_\odot$, $r_g = (4.02 \pm 0.04)$ kpc and $\gamma = 0$.

\section{CONCLUDING REMARKS}
In the following we summarize our results.
\begin{itemize}
\item The antigravitational effect of DE is well visible on the galaxies that lie belonging the region delimited by the ZGR, distance at which the total force acting on the galaxy cluster is null. 
\item We reproduced four Hubble diagrams of the cluster, finding that in both the models in which the DE is switched on (S2 and S4), galaxies out of the ZGR have a radial velocity that increases, with a linear trend, respect to the distance from the cluster centre.
\item We followed the merging among galaxies leading to the formation of substructures with mass M$_{\rm cen}\simeq$ 3.6$\%$ M$_{cl}$ and a density profile well described by a Dehnen profile ($\gamma = 0$). 
\end{itemize} 
These results allow us to conclude that the LHF can be ascribed to a local effect of DE also in a volume of few megaparsecs.

\bibliographystyle{ws-procs975x65}

\bibliography{biblio}

\end{document}